\documentclass[a4paper,11pt]{article}
\pdfoutput=1 
\usepackage{jcappub} 
\usepackage[T1]{fontenc} 
\usepackage{subcaption}
\usepackage{stmaryrd}

\title{\boldmath SNIa detection in the SNLS photometric analysis using Morphological Component Analysis}

\author[a,b,1]{A. M{\"o}ller,\note{Corresponding author.}}
\author[a]{V. Ruhlmann-Kleider,}
\author[c]{F. Lanusse,}
\author[a]{J. Neveu}
\author[a]{N. Palanque-Delabrouille}
\author[c]{and J.-L. Starck}

\affiliation[a]{Irfu, SPP, CEA Saclay, \\ F-91191 Gif sur Yvette cedex, France.}
\affiliation[b]{Universit{\'e} Paris Diderot - Paris 7,\\ 75013 Paris, France.}
\affiliation[c]{Laboratoire AIM, UMR CEA-CNRS-Paris 7, Irfu, SAp, CEA Saclay, \\ F-91191 Gif sur Yvette cedex, France.}

\emailAdd{anais.moller@cea.fr}
\emailAdd{vanina.ruhlmann-kleider@cea.fr}
\emailAdd{francois.lanusse@cea.fr}
\emailAdd{jeremy.neveu@cea.fr}
\emailAdd{nathalie.palanque-delabrouille@cea.fr}
\emailAdd{jstarck@cea.fr}

\abstract{Detection of supernovae (SNe) and, more generally, of transient events in large surveys can provide numerous false detections. In the case of a deferred processing of survey images, this implies reconstructing complete light curves for all detections, requiring sizable processing time and resources. Optimizing the detection of transient events is thus an important issue for both present and future surveys. We present here the optimization done in the SuperNova Legacy Survey (SNLS) for the 5-year data deferred photometric analysis.
In this analysis, detections are derived from stacks of subtracted images with one stack per lunation. The 3-year analysis provided 300,000 detections dominated by signals of bright objects that were not perfectly subtracted. Allowing these artifacts to be detected leads not only to a waste of resources but also to possible signal coordinate contamination.
We developed a subtracted image stack treatment to reduce the number of non SN-like events using morphological component analysis. This technique exploits the morphological diversity of objects to be detected to extract the signal of interest. At the level of our subtraction stacks, SN-like events are rather circular objects while most spurious detections exhibit different shapes. A two-step procedure was necessary to have a proper evaluation of the noise in the subtracted image stacks and thus a reliable signal extraction. We also set up a new detection strategy to obtain coordinates with good resolution for the extracted signal. SNIa Monte-Carlo (MC) generated images were used to study detection efficiency and coordinate resolution.
When tested on SNLS 3-year data this procedure decreases the number of detections by a factor of two, while losing only  10$\%$ of SN-like events, almost all faint ones. MC results show that SNIa detection efficiency is equivalent to that of the original method for bright events, while the coordinate resolution is improved.}

\begin{document}
\maketitle
\flushbottom

\section{Introduction}

Surveys of distant type Ia supernovae (SNe Ia) revealed at the end of the twentieth century the acceleration of the expansion of the Universe \citep{Riess:1998uy,Perlmutter:1999tu}. Since then, other surveys such as SNLS and SDSS-II \citep{Betoule:2014ui} have been set in place to obtain measurements of SNe Ia with higher precision. The first step for detecting SNe events is to make a sample of transient events to be later classified. Detection using only photometry with difference images in one filter, where a reference image is subtracted, provides a good approach. However, difference images are filled with various artifacts  from instrumental defects and incomplete subtraction of permanent objects. Disentangling real transient events from artifacts becomes an important requirement especially for photometric only pipelines such as the one developed in the deferred analysis of SNLS \citep{Bazin:2011em}. This is also of interest for future surveys which will process large amounts of data, such as LSST which expects to detect one million SNe per year \citep{Abell:2009aa}. 
 
SNLS is part of the Deep Synoptic Survey conducted on the Canada-France-Hawaii Telescope (CFHT). It was designed for detecting hundreds of SNe Ia in a redshift range between $0.2$ and $1$. Using the MegaCam imager \citep{boulade}, an array of 36 CCD with 340 million of pixels, four one square degree fields were targeted throughout 5 to 7 consecutive lunations per year using four different broadband filters $g_M$, $r_M$, $i_M$ and $z_M$ in the wavelength range from $400$ to $1000$ $nm$. Images were preprocessed at CFHT to perform flat-fielding and to remove defects. These pre-processed images were submitted to different pipelines. The real-time one provides detections of SNIa candidates and includes the result of spectroscopic follow-up for further classification and redshift determination \citep{Astier:2005jb}. This pipeline will not be addressed in this work. 

The deferred photometric pipeline is independent of this real time analysis. Transient events are detected in one filter and multi-band light-curves are processed for all detections. Then, these light-curves are used to select SN-like events which are assigned host galaxy photometric redshifts from an external catalogue. Light-curves and redshifts are then used to classify objects in SN Ia and core collapse types. The feasibility of detecting SNIa with this deferred analysis was proven for the 3-year SNLS data in \cite{Bazin:2011em}. In the era of large future surveys, spectroscopic resources will be limited for candidate follow-up and classification, which makes photometric pipelines interesting to study, e.g. \cite{Campbell:2013ho}.

The SNLS photometric pipeline is described in more detail in \cite{Bazin:2011em}. In the following we will summarize the main features of the detection step which are relevant to our study. Detection of transient events is done only in the $i_M$ filter because distant SNe in SNLS have their maximum flux in this band. Reference images are constructed for each field from a set of best quality images which are coadded. Each image of the survey has the reference image subtracted. The subtraction is done using determination of the sky background and a convolution kernel which allows the subtraction to be adapted to different observing conditions. In order to increase the signal-to-noise ratio, subtracted images for each lunation are stacked. Lunation detection catalogues are constructed from these subtracted image stacks keeping only events which have a S/N ratio of $2.5$ $\sigma$ w.r.t. the sky background. The final detection catalogue is obtained by merging all lunation catalogues. All detection catalogues are constructed using SExtractor \citep{SExtractor}.

\begin{figure}[htbp]
\centering
\begin{subfigure}{.3\linewidth}
\centering
\includegraphics[height=4cm]{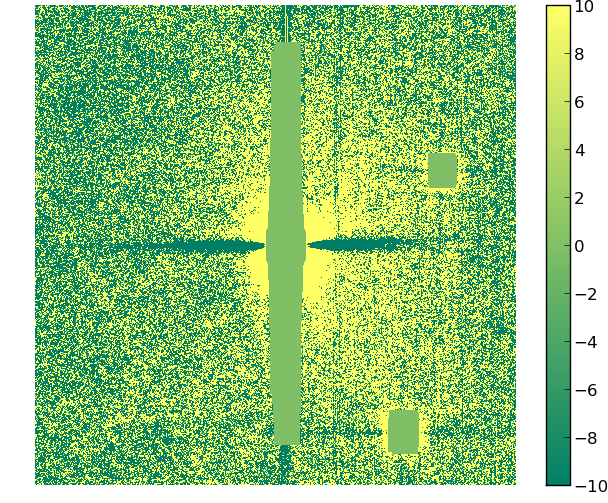}
\caption{}\label{subfig:saturated}
\end{subfigure}
\hfill
\begin{subfigure}{.3\linewidth}
\centering
\includegraphics[height=4cm]{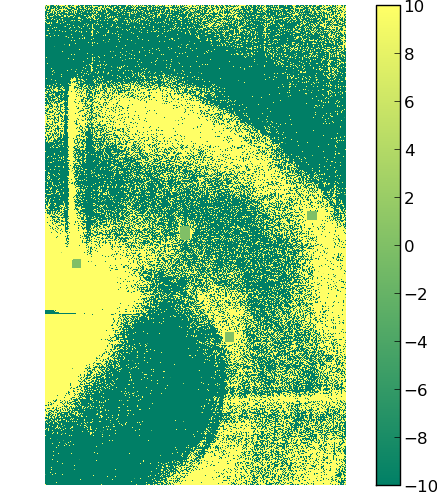}
\caption{}\label{subfig:mounting}
\end{subfigure}
\hfill
\begin{subfigure}{.3\linewidth}
\centering
\includegraphics[height=4cm]{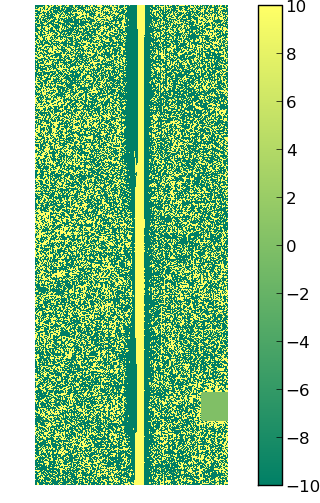}
\caption{}\label{subfig:sampling}
\end{subfigure}
\caption{Different defects on the subtracted image stacks that yield spurious detections on large scale: (a) shows a saturated star with some areas masked by subtraction, (b) a saturated star plus the camera mounting shadow and (c) defects from sampling and dead pixel lines. }\label{fig:defects}
\end{figure}

\begin{figure}[tbp]
\begin{subfigure}{.45\linewidth}
\centering
\includegraphics[height=3.5cm]{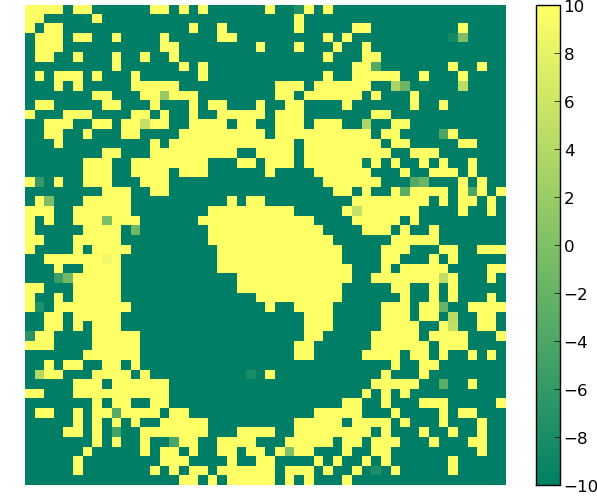}
\caption{}\label{subfig:dipole}
\end{subfigure}
\begin{subfigure}{.45\linewidth}
\centering
\includegraphics[height=3.5cm]{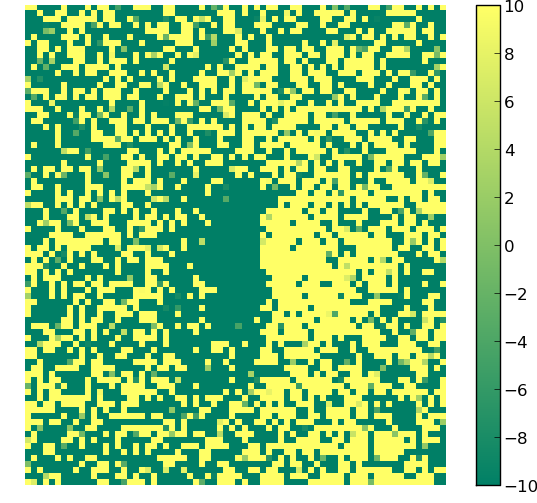}
\caption{}\label{subfig:dipole2}
\end{subfigure}
\caption{Defects on the subtracted image stacks that yield spurious detections on small scale: (a) and (b) dipoles from imperfect galaxy subtraction. These are adjacent positive and negative areas on the stacks.}\label{fig:defects2}
\end{figure}

For the SNLS 3-year (SNLS3) analysis \citep{Bazin:2011em}, the detection resulted in 300,000 transient candidates for which four-filter light curves were reconstructed. However, detections were dominated by spurious objects due to bad subtraction. Spurious detections came mostly from imperfectly subtracted objects such as bright stars, resampling defects and masks (see e.g. Figures \ref{fig:defects}, \ref{fig:defects2}). Processing light curves of such a large number of detections knowing that $80\%$ of those will be rejected by the early steps of the scientific analysis and do not contribute to science results \citep{Bazin:2011em} represents a waste of time. Therefore, in order to reduce the number of detections, it is necessary to disentangle true signal from artifacts in subtracted image stacks. 

In this paper we present a new approach for improving transient event detection based on morphological component analysis \citep{Starck:2004ur} for difference image stacks in the SNLS deferred processing. Our goal is to obtain a reduction of the number of detections while limiting the loss of SNe Ia in the detection sample. We exploit the different morphologies of objects in the stacks to separate transient objects from artifacts. Other methods to achieve such a goal exist, such as the one recently introduced by \cite{duBuisson:2014wz} based on machine learning and principal component analysis using SDSS images. We also describe a new strategy for extracting signal coordinates from our cleaned image stacks with a good resolution.

The outline of the paper is as follows. Morphological component analysis is introduced in Section 2. The method proposed to clean subtracted image stacks in order to severe spurious detections is presented in Section 3. Section 4 describes the new detection strategy based on the cleaned stacks. Results on SNLS3 data and MC efficiency and coordinate resolution studies are presented in Section 5.

\section{Morphological Component Analysis}
The Morphological Component Analysis (MCA) framework assumes that an observed image can be described as the sum of several components, each exhibiting a distinct morphology. The aim of MCA is to leverage these characteristic morphologies to disentangle the different components of an image. More formally, given an image $X$, assumed to be the sum of $K$ morphologically different components $x_k$,
\begin{equation}
\label{eq:sum_morpho}
	X = \sum_{k = 1}^{K} x_k \;, 
\end{equation}
morphological component analysis can be used to recover each individual contribution $x_k$. In the context of SN detection in image stacks, this approach can be used to disentangle transient objects from artifacts as they exhibit different morphologies.

To actually perform this separation, MCA relies on the theory of sparse representation of signals.  Any signal can be represented in a number of different domains (e.g. time domain, Fourier domain, wavelet domain) but the coefficients of this signal will exhibit different properties depending on the domain. One property of particular interest is the so called sparsity of the coefficients, i.e. the property that only a small number of coefficients are non zero. As a general rule, the coefficients of the signal will be sparse when the basis functions of the domain (so called \textit{atoms}) are very similar to the signal itself. 

In fact, this sparsity property is extremely desirable as it can be used as a very powerful prior in the regularization of a wide range of inverse problems. Some applications to astronomy and astrophysics include denoising \citep{Beckouche:2013}, deconvolution \citep{Schmitt:2012}, blind source separation for CMB analysis \citep{Bobin:2013}, weak gravitational lensing \citep{Leonard:2014}.

More formally, let us denote $\alpha$ the coefficients of a signal $x$ in a \textit{dictionary} $\mathbf{\Phi}$ (a dictionary being the set of atoms $\phi_i$ of a given domain):
\begin{equation}\label{eq:scales}
	x = \mathbf{\Phi} \boldsymbol{\alpha} = \sum_i \phi_i \alpha_i \;,
\end{equation} 
If $x$ is sparse in dictionary $\mathbf{\Phi}$ then only a small number of coefficients in $\boldsymbol{\alpha}$ are non zero. Given image $X$ defined in \eqref{eq:sum_morpho} as the sum of $K$ different morphological components, let us introduce $K$ different dictionaries $\mathbf{\Phi}_k$, each adapted only to the particular morphology of component $x_k$ i.e. such that the $\boldsymbol{\alpha}_k$ coefficients, $\{ \alpha_{ki}\}$, of $x_k$ in $\mathbf{\Phi}_k$ are sparse but not the coefficients of $x_l$ for $l \neq k$. Then performing the separation of the different morphological components can be achieved by finding an optimal set of coefficients $\boldsymbol{\alpha}_k$ maximizing the sparsity of the decomposition of each component in the corresponding dictionary. 

The Morphological Component Analysis (MCA) algorithm has been proposed by \cite{Starck:2005} as a practical way to perform this decomposition as the solution of an $\ell_1$ minimisation problem, where $\ell_1$ denotes the $\ell_1$-norm. Observed images, $Y$, are assumed to be a combination of signals, $X$, plus some noise, $N$. The decomposition algorithm solves iteratively the following optimization problem:
\begin{equation}\label{eq:norm}
\min_{\boldsymbol{x}_1, \ldots,  \boldsymbol{x}_K} \sum_{k=1}^{K} \parallel \mathbf{\Phi}^{*} \boldsymbol{x}_k \parallel_1  \quad such \, that \quad \parallel Y - \sum_{k=1}^K \boldsymbol{x}_k \parallel_2 \leq \sigma \;,
\end{equation} 
where $\mathbf{\Phi}^{*} \boldsymbol{x}_k = \boldsymbol{\alpha_k}$ and $\sigma$ is the standard deviation of the noise contaminating the data, assumed to be stationary and Gaussian distributed. Note that dictionaries and scales are chosen prior to minimization. The $\ell_1$-norm promotes the sparsity of the decomposition of each component \citep{Starck:2010}. At convergence, each morphological component is obtained as $x_k = \mathbf{\Phi}_k \boldsymbol{\alpha}_k$. This component reconstruction can be restricted to a sub-sample of $\{\alpha_{ki}\}$, for example to some size scales in a given dictionary.


\section{Reducing artifacts in SNLS subtraction stacks}

Morphological component analysis allows to disentangle artifacts from other signals and can be adapted to treat subtracted image stacks in SNLS. First, we choose dictionaries which characterize signal and artifacts distinctively at different size scales. Then, we present a two-step treatment designed to extract interesting SN-like signals and rejects spurious detections. 

A sub-sample of SNLS3 data was used to characterize artifacts and to assign the algorithm parameter values. Field D4 was chosen since it is a summer field with very good observing conditions and a large number of both detections and events classified as SN-like candidates as in \cite{Bazin:2011em}.

\subsection{Choice of dictionaries}\label{subsection:dictionaries}

The aim of the filtering approach presented in this section is to separate the signal of interest (SN-like events) from a complex background. The latter is constituted by noise, defects that cannot be subtracted (e.g. Figure~\ref{fig:defects}) and features from imperfect subtractions (e.g. Figure~\ref{fig:defects2}).

Because these artifacts are structured, a naive strategy based on detection through a simple threshold on signal-over-noise ratio yields a large number of spurious detections. Our aim is to leverage additional morphological information to separate the signal of interest from artifacts and noise, by exploiting their stark contrast in both shape and scale.

As explained in the previous section, the MCA algorithm separates images into a number of morphological components, using the sparsity level of each component in appropriate dictionaries as a discriminant. Therefore, in the case of the SNLS data, it is important to select, on one hand, a dictionary adapted to the morphology of the signal of interest and, on the other hand, additional dictionaries adapted to the artifacts we want to reject. More information on available dictionaries can be found in \citep{Starck:book}.

\begin{figure}
\begin{subfigure}{.22\linewidth}
\centering
\includegraphics[height=3.5cm]{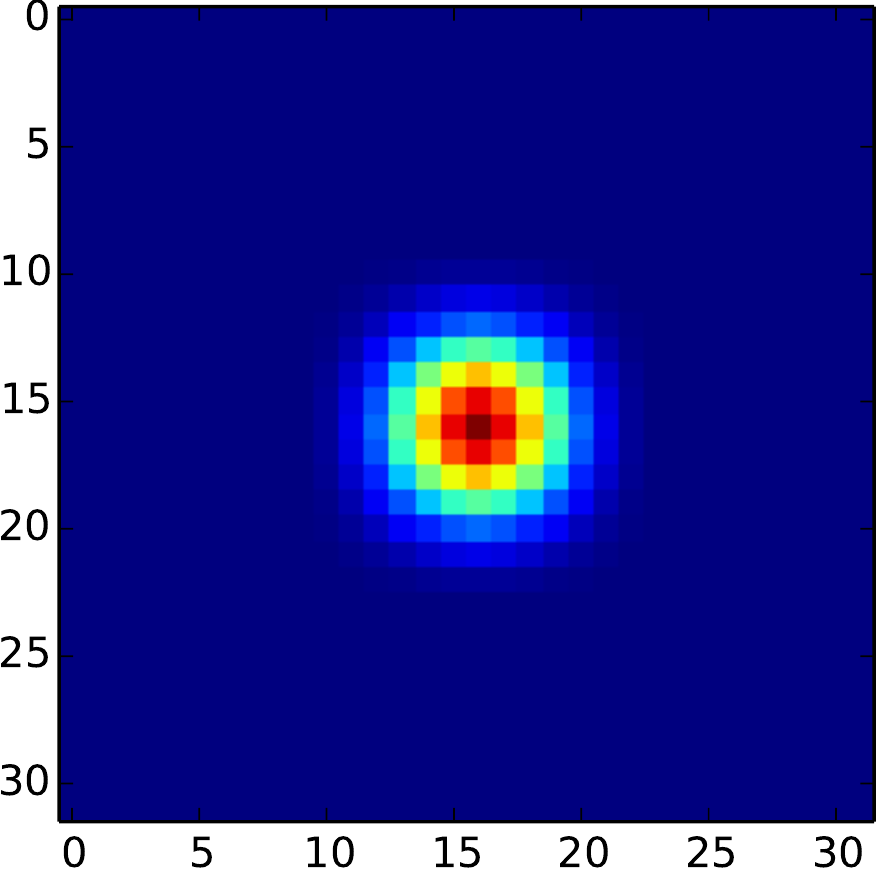}
\caption{Starlet}
\label{fig:starlet_atom}
\end{subfigure}
\hfill
\begin{subfigure}{.22\linewidth}
\centering
\includegraphics[height=3.5cm]{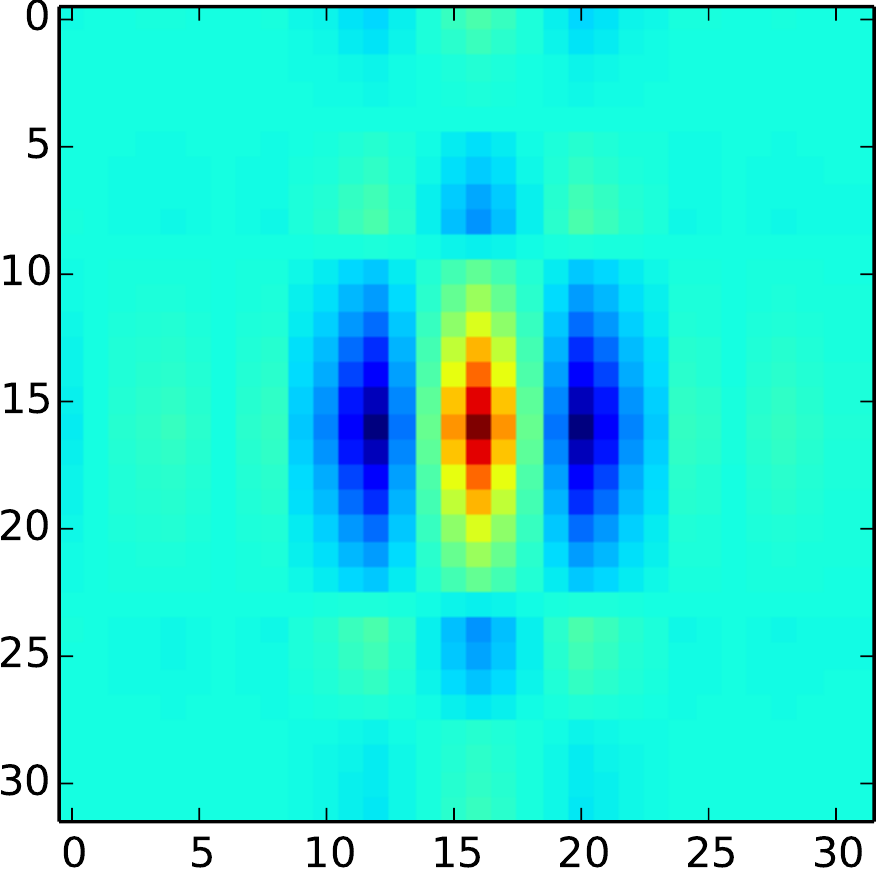}
\caption{Bi-orthogonal}
\label{fig:bi-orthogonal}
\end{subfigure}
\hfill
\begin{subfigure}{.22\linewidth}
\centering
\includegraphics[height=3.5cm]{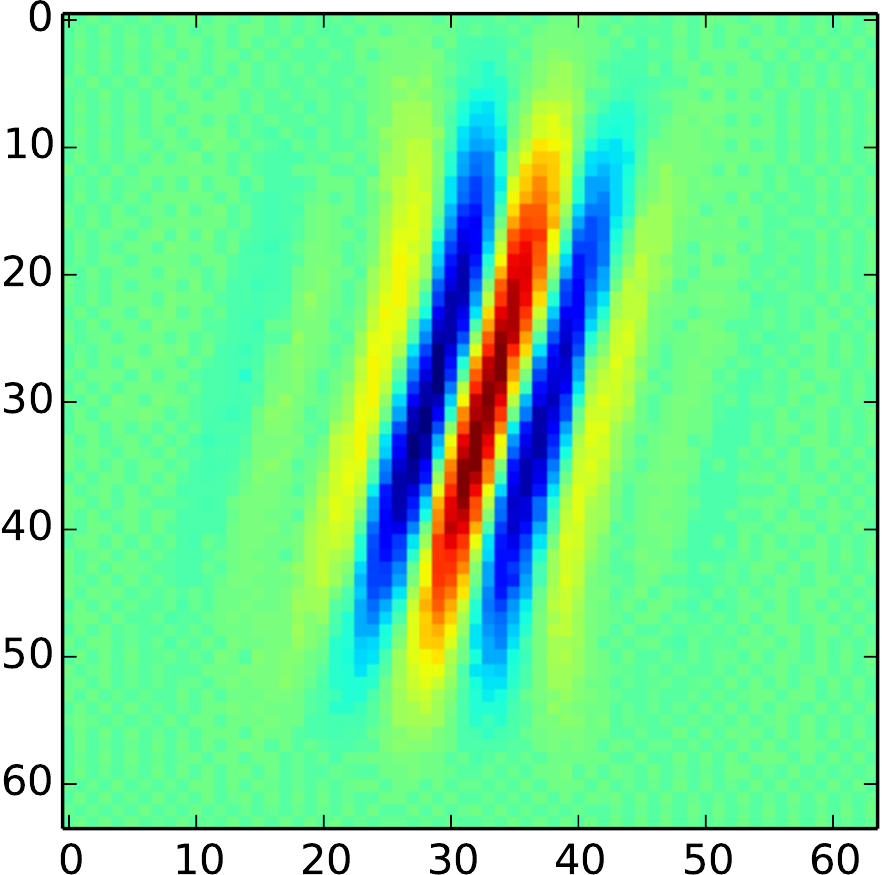}
\caption{Curvelet}
\label{fig:curvelet}
\end{subfigure}
\hfill
\begin{subfigure}{.22\linewidth}
\centering
\includegraphics[height=3.5cm]{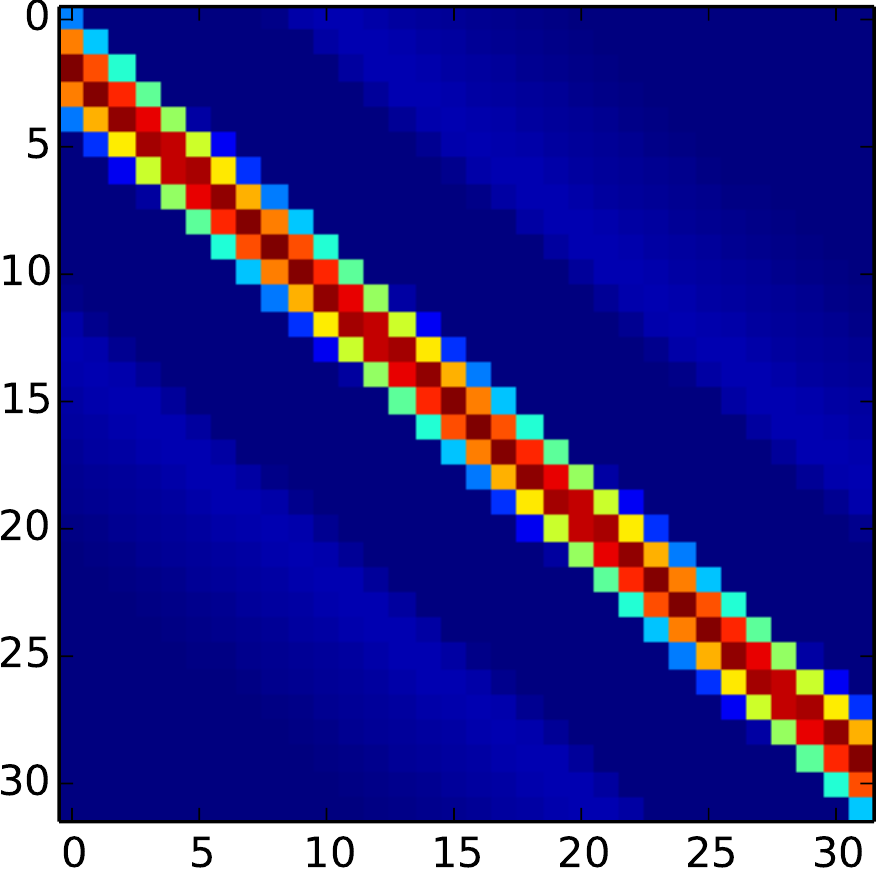}
\caption{Ridgelet}
\label{fig:ridgelet}
\end{subfigure}
\caption{Typical atoms from the dictionaries used in the MCA algorithm. (a) starlet atom representing circular-like signals, (b) bi-orthogonal wavelets for dipole features, (c) curvelets for elliptical signals and (d) ridgelets representing line features.}
\end{figure}

SN-like signals are small scale circular type shaped objects. A wavelet based dictionary is suited to this kind of morphology. We choose in particular the starlet dictionary since it is composed of isotropic atoms, especially efficient for representing positive structures such as our SN candidates. An example of a starlet atom is presented on Figure~\ref{fig:starlet_atom}.

For the small scale artifacts presented in Figure~\ref{fig:defects2} we adopt a bi-orthogonal wavelet dictionary (Figure \ref{fig:bi-orthogonal}). These artifacts result from improper subtraction of galaxies which lead to characteristic dipole features. The bi-orthogonal dictionary has the advantage of representing such features more efficiently than the starlet, enabling us to discriminate these artifacts from the signal.

For large scale curved or line artifacts such as the one in Figure~\ref{fig:defects}, we adopt curvelet and ridgelet dictionaries. The curvelet dictionary is composed of localized, elongated atoms, at different scales, which are known to provide a sparse representation for curved features, see Figure \ref{fig:curvelet}. The ridgelet atoms are line of different widths and orientations (see Figure \ref{fig:ridgelet}) which are perfect to represent the second type of artifacts.

An important aspect of all the dictionaries introduced here is that they are based on multiresolution transforms, meaning that they can be used to probe features at different discrete scales $j \in \llbracket 0, N \rrbracket$. Typically, atoms of these transforms at scale $j$ have a characteristic size of $2^j$, starting with the finest resolution with details at the pixel scale for $j=0$. Note that this $j$ scale index is embedded together with pixel coordinates in the $i$ index in equation \ref{eq:scales}. The advantage of choosing different scales for each dictionary is that we are able to separate small scale signals from large scale defects. We make use of this scale information within the MCA algorithm as explained in the next section.

\subsection{First treatment: removal of main artifacts}
\begin{figure}[htbp]
\begin{subfigure}{.22\linewidth}
\centering
\includegraphics[height=3cm]{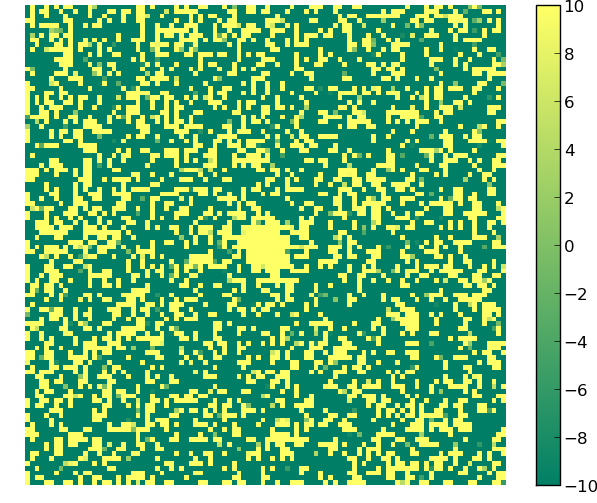}
\caption{}
\end{subfigure}
\hfill
\begin{subfigure}{.22\linewidth}
\centering
\includegraphics[height=3cm]{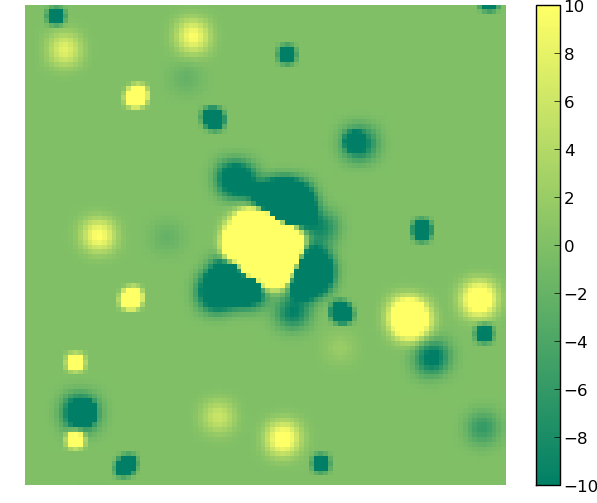}
\caption{}
\end{subfigure}
\hfill
\begin{subfigure}{.22\linewidth}
\centering
\includegraphics[height=3cm]{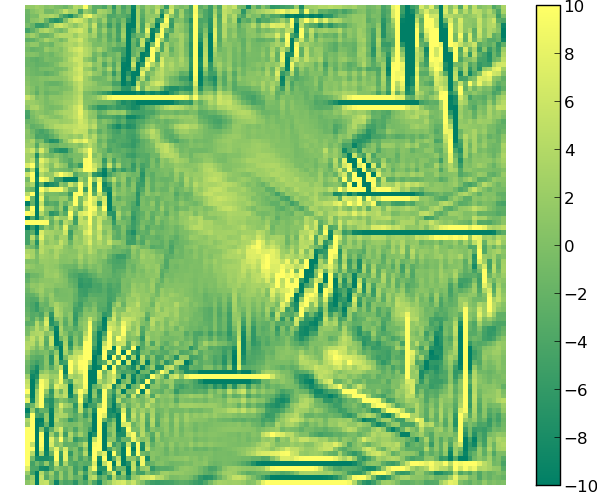}
\caption{}
\end{subfigure}
\hfill
\begin{subfigure}{.22\linewidth}
\centering
\includegraphics[height=3cm]{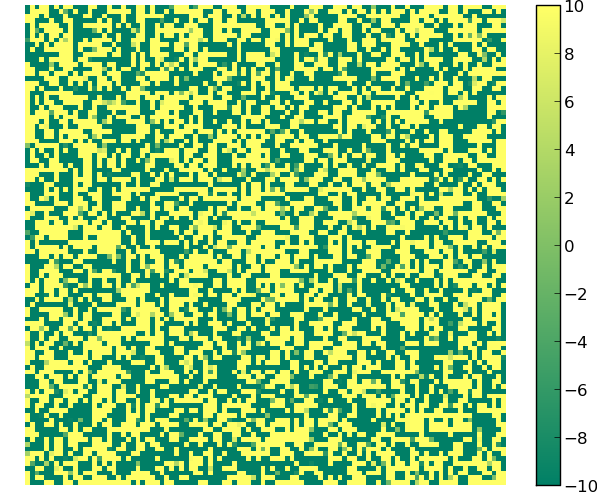}
\caption{}
\end{subfigure}
\hfill
\caption{MCA Decomposition of a SNIa event. (a) shows the original subtracted image stack centered on the SN event (yellow spot), (b) the starlet component, where the SN signal (yellow spot) is surrounded by remaining galaxy residuals (green spots), (c) the curvelet component and (d) the residuals left after the decomposition.}\label{fig:sn_decomposition}
\end{figure}

The MCA algorithm in \cite{Starck:2005} was adapted to treat our subtracted image stacks. To disentangle signal artifacts we chose the previously described dictionaries: starlets, bi-orthogonal wavelets, curvelets and ridgelets. SNLS3 D4 test sample was used to decompose known artifacts and select the scales in each dictionary that allowed the best characterization. The best choices were 5 scales in dictionaries representing mostly artifacts (curvelets, ridgelets, bi-orthogonal), and 3 scales in the starlet dictionary which efficiently decompose circular-like signals as can be seen in Figure \ref{fig:sn_decomposition}. Both positive and negative signals were decomposed since some SNe may have part of their flux included in the references, in which case subtraction yields a negative signal.

\begin{figure}[htbp]
\centering
\begin{subfigure}{.3\linewidth}
\centering
\includegraphics[height=3cm]{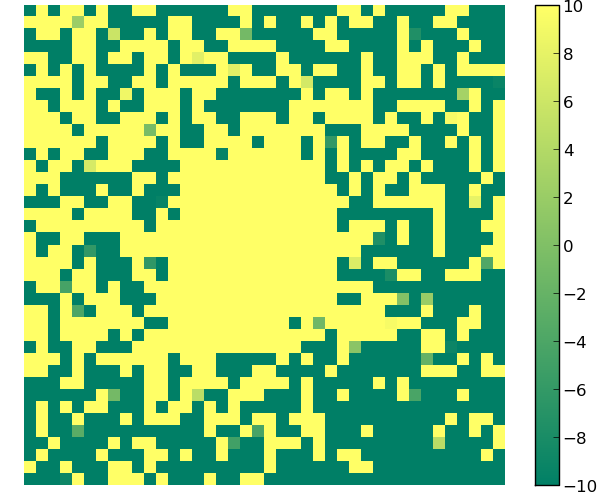}
\caption{}
\end{subfigure}
\hfill
\begin{subfigure}{.3\linewidth}
\centering
\includegraphics[height=3cm]{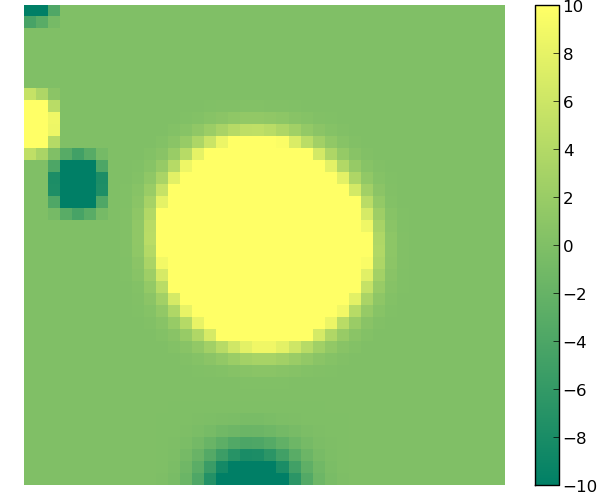}
\caption{}
\end{subfigure}
\hfill
\begin{subfigure}{.3\linewidth}
\centering
\includegraphics[height=3cm]{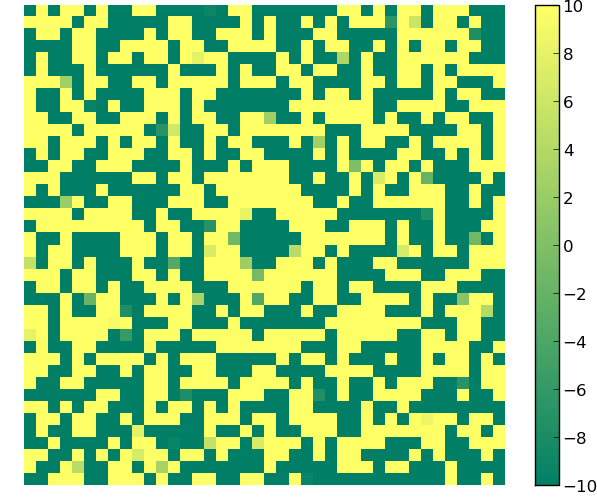}
\caption{}
\end{subfigure}
\caption{MCA Decomposition of a SNIa event where some part of the signal leaks into residuals. (a) shows the original subtracted image stack,   (b) the starlet component and (c) the residuals after decomposition containing part of the signal. }\label{fig:sn_leak}
\end{figure}

Once dictionaries and scales are set, other algorithm parameters must be chosen such as the number of iterations in the optimization.
The choice of parameters resulted from a trade-off between reducing the total number of detections and keeping most of the SN-like objects in the D4 test sample. A compromise between number of iterations and computation time was achieved with 30 iterations for the decomposition. 

The transforms used in the algorithm, especially that of the curvelet dictionary, do not scale well with the image size and too much CPU time and memory would be required for a SNLS survey image of 2176 by 4912 pixels. Therefore, tiling of the images was done both to reduce time and memory resources and to allow parallel processing. For reference, one SNLS image divided in 8 tiles requires on average 6 days of HS06 CPU time and 500 Mb of virtual memory to be treated.

The algorithm assumes a stationary and Gaussian noise in the input images which is not the case for our subtracted image stacks, that are built from the coaddition of subtracted images spanning several weeks of observations. Thus, the signal we aimed at recovering was not properly decomposed and was partially in the residuals, e.g. Figure \ref{fig:sn_leak}. To tackle this, a second treatment was developed which uses as input the starlet component and the residuals of our first treatment decomposition (e.g. components (b) and (d) of Figure \ref{fig:sn_decomposition}).

\subsection{Second treatment: signal extraction with varying noise }

A utility based on the algorithm in \cite{Starck:book} was developed. It handles non-stationary noise and exploits further morphological decomposition. Non-stationary noise requires varying the threshold in the decomposition depending on the position of the analyzed pixel. Such a feature can be easily implemented in the wavelet dictionary since it can handle actual noise maps. The latter were computed from the first treatment output images using a median absolute deviation estimator. This computation used a sliding window with a fixed size larger than what is expected for a SN-like signal. Since some of the SNe may have part of their flux included in the references, both positive and negative signals were treated.

\begin{figure}[tbp]
\begin{subfigure}{1\linewidth}
\centering
\includegraphics[width=1\textwidth]{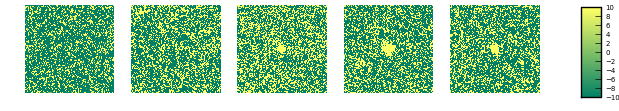}
\caption{}
\end{subfigure}
\begin{subfigure}{1\linewidth}
\centering
\includegraphics[width=1\textwidth]{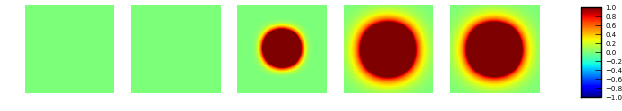}
\caption{}
\end{subfigure}
\caption{A SNIa event shown in different lunations around maximum light in the original subtracted image stack (a) and after both treatments of the cleaning procedure (b).}\label{fig:sn_sts_snd}
\end{figure}

We used again the starlet dictionary but this time handling varying noise, in order to select significant coefficients. All signals present in the output can be considered as morphologically compatible with circular-like objects. An example can be seen in Figure \ref{fig:sn_sts_snd}. 

This utility does not require tiling images since only the wavelet dictionary is used. One SNLS image takes on average 3 hours of HS06 CPU time and 100 Mb of virtual memory to be treated.

\section{New Detection strategy}
A detection strategy includes both extracting events from an image and reconstructing their coordinates. Event extraction depends on the image and its characteristics, e.g. its local noise information.  The TERAPIX tool SExtractor \citep{SExtractor} was used for the whole detection strategy both in the original procedure and the new procedure adjusting its parameters accordingly.

\begin{figure}[tbp]
\begin{center}
\includegraphics[width=\textwidth]{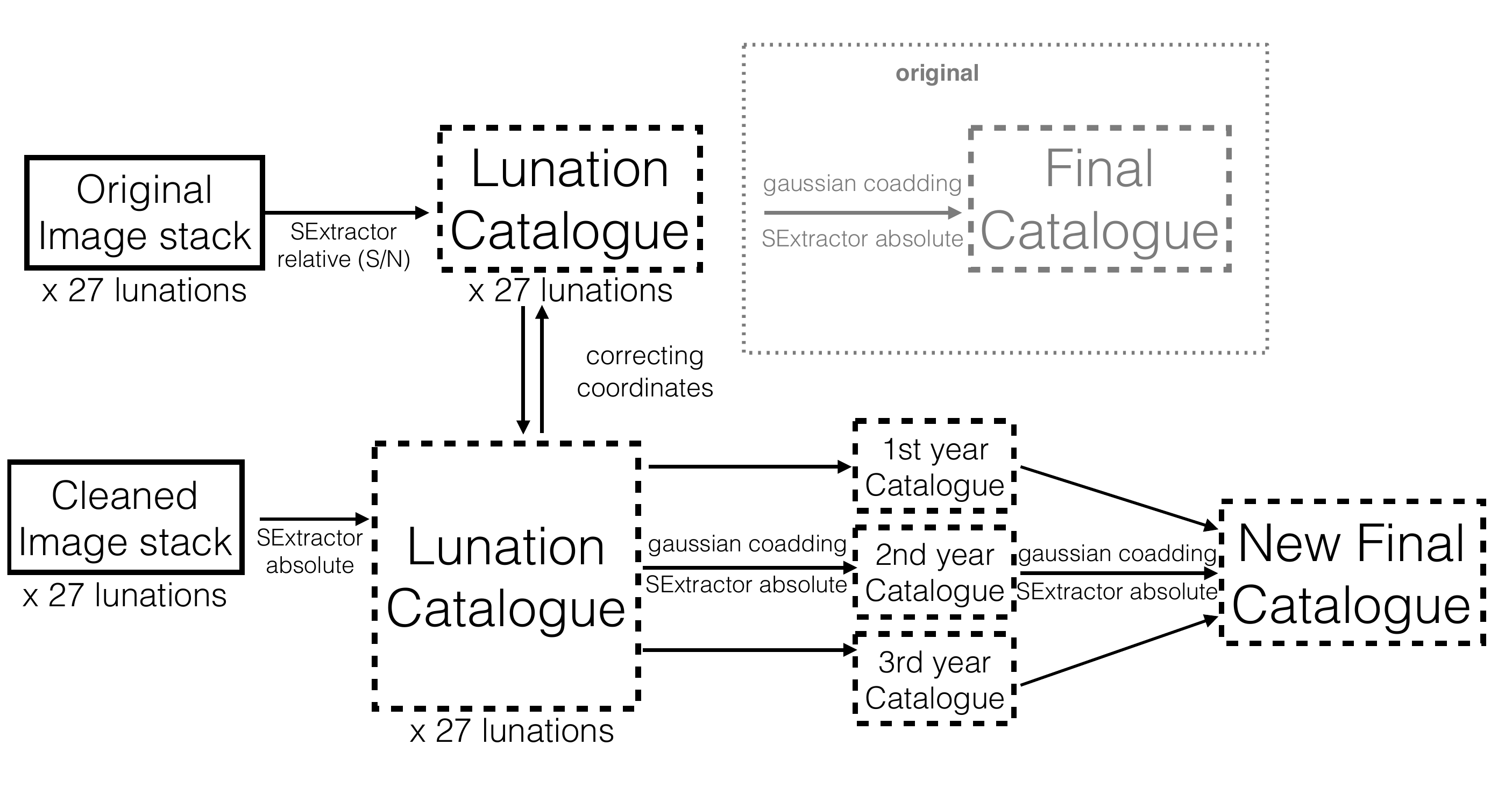}
\end{center}
\caption{New detection strategy schema. Doted lines represent catalogue ASCII files while continuous lines images.}\label{fig:detection_schema}
\end{figure}

\begin{figure}[tbp]
\begin{subfigure}{\linewidth}
\centering
\includegraphics[width=\textwidth]{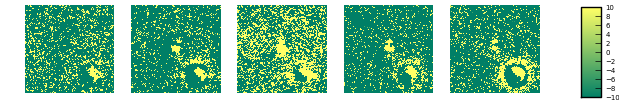}
\caption{}\label{fig:gal_residuals_original}
\end{subfigure}
\begin{subfigure}{\linewidth}
\centering
\includegraphics[width=\textwidth]{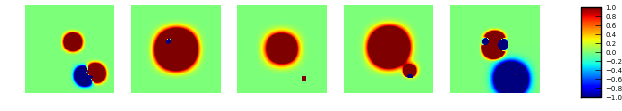}
\caption{}\label{fig:gal_residuals_cleaned}
\end{subfigure}
\caption{A SNIa event (center of the image) with galaxy residuals shown in different lunations around maximum light in the original subtracted image stacks (a) and after cleaning (b). In the original stacks, galaxy residuals are present in all lunations. The cleaning removes them in some cases.}\label{fig:gal_residuals}
\end{figure}

In \cite{Bazin:2011em} the detection strategy consisted in constructing lunation catalogues with SExtractor with deblending, requiring for each detection at least 4 pixels with signal of more than $2.5\sigma$ w.r.t. sky background. A final detection catalogue was obtained by merging all lunation catalogues obtained in three years and converting the result into an image where each detection was replaced by a Gaussian of height and width of $1$. This image was processed with SExtractor selecting only pixels with a content above a value of $0.01$ and deblending objects.  In this way, any object detected on several lunations at the same position (within a pixel) gave only one detection, with a position averaged over all lunation stacks. This is described on the top part of Figure \ref{fig:detection_schema}. Even though the lunation stacking reduces the number of detections (typically by a factor 3), when adding many years it can degrade signal coordinate resolution due to close-by spurious detections as can be seen in Figure \ref{fig:gal_residuals_original}. 

Our two-step treatment outputs image stacks which do not have the same properties as the original subtracted image stacks. The noise has been removed and sources are reconstructed using inverse transformations. As transformed images they have less objects but coordinates cannot be extracted accurately from them. We thus propose a new detection strategy (see Figure \ref{fig:detection_schema} bottom) which also addresses the degradation of coordinate resolution when using several years of data. This strategy was set up on SNLS3 D4 data.

Lunation catalogs are constructed from our cleaned subtraction stacks using SExtractor, requiring at least 200 pixels with a signal value above one to confirm an object. Deblending is imposed in order to separate adjacent objects. The values of the SExtractor parameters were tuned using the SNLS3 test sample. They resulted from a trade-off between the reduction of the total number of detections and the number of SN-like objects detected on the cleaned image stacks. To each object detected in a lunation we assign the coordinates of the closest detection in the same lunation catalogue of the original procedure. In this way we maintain the reduced number of candidates while having precise coordinates.

In the original procedure coordinates were averaged over all lunations which degrades signal coordinate resolution due to close-by spurious detections. The latter are not always completely removed by cleaning as can be seen in Figure \ref{fig:gal_residuals_cleaned}. When adding data from other seasons, the coordinate resolution degradation becomes even more important. Real SN-like events can be present in at most three adjacent lunation catalogs but not over several seasons. Hence, to address this in the new procedure we first build a catalogue for each season as we did for the final catalogue in the original procedure. Then, we build the new final catalogue from the season catalogues in the same way. 

In this way, coordinate averaging is done first for a season were a transient object can be present and then detections are added from other years. It is equivalent to assigning a weight for a given detection taking into account that a SN will be detected only during one season.

\section{Results}

\subsection{SNLS data}
The SNLS3 D4 test sample contained 90,971 detections from which 362 events were extracted as SN-like objects as described in \cite{Bazin:2011em}. After our processing, the number of detections is reduced to 40,575. This represents more than a factor 2 reduction on the number of candidates to be further processed. Loss of SN-like candidates is less than $5\%$ and all lost events are faint (observed magnitude at peak in $i_M > 24.2$) and so not suitable for further cosmological analysis. 
The complete procedure with parameters as determined from D4 data was then applied to the three other fields. Results are summarized in Table \ref{table:fields_det}. The reduction of the number of detections is similar in all fields. The loss of SN-like events is less than 5$\%$ in D3 and 15$\%$ in D1 and D2. It must be noted that D1 and D2 are the fall and winter fields which have less suitable weather conditions than D3 and D4. All lost events are faint with the exception of one medium brightness event in D1 which is lost during our new detection procedure. This event is found on the output images of the two-step treatment but the number of pixels above threshold is smaller than our criteria to validate a detection. 

\begin{table}[tbt]  
\centering
\begin{tabular}{|c| c | c | c | c|}
\hline
    	 & \multicolumn{2}{c}{Old procedure} & \multicolumn{2}{|c|}{New procedure} \\ 
    	 \hline
    	Field & $\#$ detected & $\#$ SN-like & $\#$ detected & $\#$ SN-like  \\  	
    	\hline
	D1 & 76,806 & 444 & 34,314 & 382 \\
	D2 & 64,763 & 300 & 28,627 & 258 \\
	D3 & 70,447 & 377 & 29,292 & 359 \\
	D4 & 90,971 & 362 & 40,575 & 346\\
	\hline
	All & 302,987 & 1,483 & 127,808 & 1,345\\
    	\hline
\end{tabular}
\caption{Number of detected and SN-like events for the original and new proceedures applied on SNLS3 data.} \label{table:fields_det}
\end{table}

\begin{table}[tbt]  
\centering
\begin{tabular}{|c| c c | c c | c c|}
\hline
	    & \multicolumn{2}{c}{\textbf{Old procedure}} & \multicolumn{4}{|c|}{\textbf{New procedure}} \\ 
    	 & \multicolumn{2}{c}{} & \multicolumn{2}{|c}{No season stacks}  & \multicolumn{2}{c|}{With season stacks} \\ 
    	 \hline
    	& coordinate & magnitude & coordinate  & magnitude  & coordinate  & magnitude  \\  
     Stack  &  resolution &  bias &  resolution &  bias &  resolution &  bias \\  		
    	 & \small$\pm 0.002 $& \small $\pm 0.0002 $& \small$\pm 0.002$ & \small $\pm 0.0002 $ & \small$\pm 0.002$ & \small $\pm 0.0002 $\\
    	\hline
    	1-year & $0.709 $ & $0.0334 $& $0.698$  & $0.0324 $  &$0.698 $ & $0.0324 $  \\
    	3-year & $0.725$ & $0.0349 $  & $0.715$& $0.0340 $ &$0.710 $ & $0.0335 $ \\
    	5-year & $0.741$  & $0.0365 $  & $0.731$ & $0.0355 $ &$0.726$ & $0.0350 $  \\ 
    	\hline
\end{tabular}
\caption{Estimated coordinate resolutions (pixels) and corresponding magnitude bias of SNIa detection original procedure, new procedure with no season stacks or complete new procedure with season stacks: for year 3 MC data (1-year stack), adding two additional years of data (3-year stack) and adding 4 additional years of data (5-year stack). Uncertainties shown here are from the statistics of generated SNe Ia.} \label{table:coord_res}
\end{table}

\subsection{Monte-Carlo efficiency and coordinate resolution}
The performance of our treatment was studied using Monte Carlo (MC) artificial images in the $i_M$ filter generated in the D1 field \citep{Ripoche}. The MC images were constructed by adding simulated supernovae to real raw images, on  host galaxies identified from deep image stacks of the CFHT-LS Deep Fields \citep{Ilbert}. 
Using a two-dimensional gaussian for modeling the galaxies, SN positions within their hosts were randomly generated up to a distance of $2\sigma$ from the host galaxy centers. This method provided compatible simulated and observed SN-host galaxy angular distance distributions. The redshift assigned to each SN was that of its host galaxy and was restricted to the range between $0.2$ and $1.2$. For each SN the $i_M$ light curve was simulated according to the SN properties (redshift, color, stretch) and the generated SN flux as deduced from the light curve at each observation date was added to the corresponding raw image. The MC images were then processed by the deferred pipeline as real images. Subtracted MC image stacks were then processed by our optimized pipeline. The corresponding results were compared to the ones of the original procedure in \cite{Bazin:2011em}. 

\begin{figure}[htbp]
 \begin{center}\includegraphics[width=9cm]{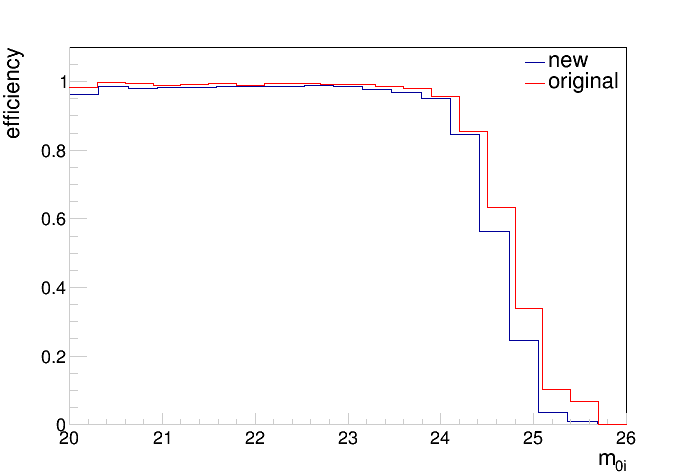}
 \end{center}
\caption{Efficiency of detection as a function of the generated peak magnitude in $i_{M}$. The new procedure (blue line) is compared to the original one (red line).}\label{fig:efficiency_only3}
\end{figure}

Detection efficiency was defined as the fraction of recovered simulated supernovae at the end of each processing. For both the original and the new procedure we computed the detection efficiency for one year of simulated SNe Ia as a function of the generated SN peak magnitude as can be seen in Figure  \ref{fig:efficiency_only3}. The efficiency is nearly magnitude independent up to $m_{0i}=23.5$ and then steeply declines at faint magnitudes. When compared to the old procedure, the new procedure corresponds to a loss of $0.5 \%$ in the plateau efficiency and a $0.2$ downward shift of the magnitude corresponding to $0.50$ efficiency. Note that after the new procedure the efficiency behavior as a function of magnitude is close to the original one and is consistent with that expected from a magnitude limited survey. This MC result will allow us to correct the Malmquist bias of the photometric sample to be derived from the new procedure in order to perform a cosmological analysis.

The SNIa coordinate resolution was studied for new and old procedures as can be seen in Table \ref{table:coord_res}. The resolution was given by the RMS of the distance between the coordinate at generation and at detection. Coordinate resolution is improved first by reducing spurious detection with our cleaning procedure (Table \ref{table:coord_res}, column 2). Further gain is obtained by the modified detection strategy (Table \ref{table:coord_res}, column 3).

The coordinate resolution for one year simulated SNe Ia in our complete optimized pipeline was found to be $0.698 \pm 0.002$ pixels to be compared to $0.709\pm0.002$ pixels in the original pipeline. We also studied the effect of adding other years of survey data (without simulated SN signal), constructing catalogues with two or 4 additional years. A degradation of coordinate resolution is seen but the new procedure handles better many years of data than the original one.

Position measurement inaccuracy leads to underestimated fluxes. Using appendix B of \cite{Guy:2010bp} we computed an indicative magnitude bias corresponding to our coordinate resolution. Thanks to the improved coordinate resolution, the new procedure applied on 5-year stacks has similar performance to the old procedure applied on 3-year stacks.The latter was found accurate enough for photometric typing as shown in \citep{Bazin:2011em}. The application of the new procedure to the whole set of SNLS data will be the subject of a future work.

\section{Discussion}

Morphological component analysis has proven to be a useful approach for cleaning subtracted image stacks such as the ones in the SNLS deferred processing.  Our experience shows that the precise nature of the input images was a key point when choosing and adapting this type of algorithms. The choice of algorithm was based on the availability of a robust tool that could decompose our subtracted image stacks efficiently and within our CPU and time resources. For adapting the algorithms to the defects present in our input images we had to use a two-step procedure. In the first step we needed several dictionaries and scales to eliminate the various artifacts. Note that many of these defects came from the fact that we used subtracted images that usually have many residuals. For the second algorithm the goal was to handle non-stationary noise (typical from stacks) in addition to SN-like signals which provided a natural choice of the starlet dictionary for the decomposition. Finally, the choice of algorithm parameters (e.g. number of iterations) was heavily dependent on the efficiency and purity we wanted to achieve and on computing resources.

We note that besides improving the subtraction algorithm itself, eliminating artifacts at the level of subtracted images instead of stacks can provide a higher reduction of the number of detections. However, this should be applied at the beginning of the survey. For implementing such methods, a thorough analysis must be done of the trade-off between gain on signal extraction and removal of artifacts with respect to the high computational and time costs of processing using dictionary decomposition.

Future surveys like LSST may detect around ten thousands SNe Ia a year \cite{Carroll:2014oja}, which is two orders of magnitude higher than in SNLS. Extrapolating what we experienced in the deferred processing of SNLS, the anticipated number of detections in LSST may be as high as $10^7$ per year which is too large to process. To reduce the number of candidate transient events to process further, cleaning images with a fast multi-resolution method can be of interest. But due to the huge number of detections, additional multi-band and temporal information will be necessary. The above arguments are valid for both real-time and deferred processings, which will both face too large numbers of detections to process. Differences between the two approaches would affect the choice of cleaning algorithms and selection criteria based on multi-band and temporal information.

\section{Conclusions}

In this paper we presented a new procedure for detecting supernovae in the SNLS photometric analysis. We developed a two-step procedure for cleaning subtracted image stacks, reducing artifacts and extracting SN-like signals using morphological component analysis. A new detection strategy, adapted to the cleaned image stacks was also presented.

The performance of the new procedure was evaluated using MC artificial images. Detection efficiency of SNeIa in the old and the new procedure is almost unchanged for bright events. However, there is a small reduction for events at higher magnitudes, which is expected since signal separation is not perfect and some SN-like signal may not be properly transformed. When applied to real SNLS3 data, 10$\%$ percent of SN-like events were lost while the number of detections was reduced by more than a factor two. Almost all lost events were faint with the exception of one medium brightness event which was lost in the detection step. This result agrees with MC findings.

Coordinate resolution of SNIa events was equivalent for one year of MC for both procedures. Furthermore, since SNLS is a five-year survey, coordinate resolution was also studied adding other years of data. The new procedure yields slightly better SNIa coordinate resolution with respect to the original procedure when adding 4 additional years of data, simulating a five-year stack. Therefore, for a five-year photometric analysis this new procedure yields a slightly smaller magnitude bias for SNe Ia when compared to the original procedure. The new procedure presented in this work will be applied to the final SNLS 5-year photometric analysis which will be the subject of a forthcoming paper.

This work is a first step on morphological component analysis applied for SN-like signal detection and may be used as a starting point for future surveys. For those surveys that will detect a large number of events such as the LSST a fast multi resolution algorithm can be of interest, provided additional information (e.g. other filters, partial light curves) is also used.

\acknowledgments

This work was done based on observations obtained with MegaPrime/MegaCam, a joint project of CFHT and CEA/IRFU, at the Canada-France-Hawaii Telescope (CFHT) which is operated by the National Research Council (NRC) of Canada, the Institut National des Science de l'Univers of the Centre National de la Recherche Scientifique (CNRS) of France, and the University of Hawaii. This work is based in part on data products produced at Terapix available at the Canadian Astronomy Data Centre as part of the Canada-France-Hawaii Telescope Legacy Survey, a collaborative project of NRC and CNRS.

\bibliographystyle{JHEP}
\bibliography{myBib}

\end{document}